\documentclass[12pt]{article} 
\usepackage{amssymb,color} 

\newcommand{\nc}{\newcommand}
\nc{\la}{\lambda} \nc{\alf}{\alpha} \nc{\La}{\Lambda} \nc{\ze}{\zeta}
\nc{\tht}{\theta} \nc{\T}{\Theta} \nc{\be}{\beta}  \nc{\eps}{\epsilon} 
\nc{\ga}{\gamma}  \nc{\De}{\Delta}  \nc{\G}{\Gamma}  \nc{\vphi}{\varphi}
\nc{\de}{\delta} \nc{\si}{\sigma}  \nc{\ka}{\kappa}   \nc{\Si}{\Sigma} 
\nc{\om}{\omega}  \nc{\qq}{\quad\quad}                \nc{\Om}{\Omega} \nc{\vrho}{\varrho}
\nc{\nf}{\infty}   \nc{\dl}{\mathop{\smash{\cal L}}}  \nc{\black}{\rule{3mm}{3mm}}
\nc{\ra}{\rightarrow}    \nc{\ol}{\overline}        \nc{\und}{\underline} 
\nc{\beq}{\begin{equation}}  \nc{\eeq}{\end{equation}}  \nc{\pt}{\partial}  
   \nc{\dst}{\displaystyle}  \nc{\na}{\nabla} 
\nc{\nnb}{\nonumber}    \nc{\bs}{\backslash}        \nc{\mb}{\mathbb}   
\nc{\sn}{{\rm sn}\,} \nc{\cn}{{\rm cn}\,}     \nc{\dn}{{\rm dn}\,} \nc{\nin}{\noindent}
\nc{\ti}{\tilde}   \nc{\wti}{\widetilde}   \nc{\h}{\hat}  \nc{\wh}{\widehat}
\nc{\tpsi}{\wti{\psi}}   \nc{\tphi}{\wti{\phi}}  \nc{\tH}{\wti{H}} \nc{\Ai}{{\rm Ai}}
\nc{\Pf}{P_{\phi}}  \nc{\Pt}{P_{\tht}}

\newcounter{muni}
\newenvironment{remunerate}{\begin{list}{{\rm \arabic{muni}.}}
{\usecounter{muni}
\setlength{\leftmargin}{0pt}\setlength{\itemindent}{38pt}}}{\end{list}}

\nc{\brm}{\begin{remunerate}}   \nc{\erm}{\end{remunerate}}

\newtheorem{nth}{Proposition}  \newtheorem{nTh}{Theorem}     

\nc{\stg}{\mathop{\smash{*}}}
\nc{\st}{\mathop{\smash{\delta}}}
\nc{\barr}{\begin{array}}   \nc{\earr}{\end{array}}   \nc{\dg}{\dagger}
\nc{\mtvb}{\mathversion{bold}}   \nc{\mtvn}{\mathversion{normal}} 

\begin{document} 

\begin{titlepage}

\hfill{7/04/2014}

\vspace{1cm}
\centerline{\Large\bf  Zoll and Tannery metrics} \vskip 0.5truecm 
\centerline{\Large\bf from a superintegrable geodesic flow}

\vskip 2.0truecm
\centerline{ \large\bf Galliano VALENT  \footnote{Sorbonne Universit\'es, UPMC Universit\'e Paris 06, UMR 7589, LPTHE, F-75005, Paris, France.\\ 
\indent\indent CNRS, UMR 7589, LPTHE, F-75005, Paris, France.\\
\indent\indent Aix-Marseille University, CNRS UMR 7332, CPT, 13288 Marseille, France.\\ 
\indent\indent Universit\'e de Toulon, CNRS UMR 7332, CPT, 83957 La Garde, France.} }
\nopagebreak

\vskip 2.5truecm

\begin{abstract} We prove that for Matveev and Shevchishin superintegrable system, with a linear and a 
cubic integral, the metrics defined on ${\mb S}^2$ and on Tannery's orbifold ${\cal T}^2$ are 
either Zoll or Tannery metrics.
\end{abstract}

\end{titlepage}

\newpage
\section{Introduction}
A family of dynamical systems on two dimensional manifolds was defined by Matveev and Shevchishin 
in \cite{ms} and shown to have a superintegrable geodesic flow. It was conjectured by these 
authors that the corresponding metrics on closed manifolds could lead to either 
Tannery or Zoll metrics, i. e. metrics for which all of the geodesics are closed. However a 
proof of this expected result was not given since the metrics were known only up to the integration 
of some first order non-linear ordinary differential equations. In \cite{vds} we were able 
on the one hand to integrate these differential equations up to the explicit form of the metrics 
in local coordinates and on the other hand to determine which metrics are globally defined 
on ${\mb S}^2$. It is the aim of this article to prove the conjecture of Matveev and Shevchishin  
and to show that it does hold also for some metrics defined on orbifolds.
 
\nin Our results are the following:
\brm
\item The surface defined on ${\mb S}^2$ is Zoll. 
\item Since in \cite{vds} only metrics defined on manifolds were of interest some of the surfaces 
were discarded because they exhibited a conical singularity, still being of finite measure. In fact Thurston introduced the concept of orbifold to deal with this new kind of geometrical objects. 
We will show that all the metrics with a conical singularity, obtained in \cite{vds}, are defined 
on Tannery's orbifold ${\cal T}^2,$ the simplest surface being known as Tannery's pear. 
\item The conserved cubic observables which bring in superintegrability describe parametrically the 
geodesic trajectories.
\erm

The structure of the article is the following: in Section 2 we present some background material, 
coming essentially from \cite{Be}, for self-containedness. In Section 3 we consider the metrics 
shown in \cite{vds} to be globally defined on ${\mb S}^2$ and prove that they are Zoll. In Section 4 
we consider Tannery's pear and in Section 5 a one parameter generalization of it for which the geodesics 
close also after two turns. In Section 6 we consider a further  generalization with two parameters which 
is Zoll. Remarkably enough these three metrics are defined on the same orbifold. In Section 7 we present 
some conclusions and prospects for open problems of interest.

\section{Basic material}
To any riemannian metric of the form
\beq
g=g_{ij}\,dx^i\,dx^j=A^2(\tht)\,d\tht^2+\sin^2\tht\,d\phi^2\qq\qq \tht\in\,(0,\pi)
\qq \phi\in\,{\mb S}^1\eeq
one associates the hamiltonian
\beq
H=\frac 12\left(\Pi^2+\frac{\Pf^2}{\sin^2\tht}\right)\qq\qq \Pi=\frac{\Pt}{A(\tht)}\eeq
Let us observe that we have two obvious conserved quantities: $H$ and $\,\Pf$ which are preserved by 
the geodesic flow and we will denote by $E$ and $L$ their values. 
The Hamilton equations give the following differential system for the geodesics
\beq
\Pi=\eps\,\sqrt{2E-\frac{L^2}{\sin^2\tht}}\qq \eps=\pm 1\qq\qq \sin^2\tht\,\frac{d\phi}{dt}=L>0.
\eeq
The choice $\,2E=1$ is quite convenient since it ensures that the $t$-coordinate is nothing but 
the arc length $s$ and we will take for initial conditions
\beq\label{init}
\tht=i\in\,(0,\pi/2)\qq \phi=0 \qq L=\sin i \qq  \Pi=0 .\eeq 
The trajectory itself is determined by the integration of
\beq
\frac{d\phi}{d\tht}=\eps\,\frac{\sin i\,A(\tht)}{\sqrt{\sin^2\tht-\sin^2 i}}.
\eeq

\vspace{3mm} 
\nin {\bf Remark:} It is important to recall that the geodesic equations are made 
out of two pieces: in the first one $\tht$ increases from $i$ to $\pi-i$ and then we have 
$\Pi\geq 0$ while in the second piece $\tht$ decreases from $\pi-i$ to $i$ and then $\Pi\leq 0$. This should be kept in mind for all the geodesic equations given below. 

The following theorem, for which a proof can be found in \cite{Be}, is of paramount importance 
to determine the closedness of the geodesics:

\begin{nTh}[Darboux] A necessary and sufficient condition in order that all the geode\-sics be closed is that 
the rotation function 
\beq\label{rotf}
R(g,i)=\int_i^{\pi-i}\frac{\sin i\,A(\tht)}{\sin\tht\,\sqrt{\sin^2\tht-\sin^2 i}}d\tht\eeq
be equal to $\,\dst\frac pq\,\pi$ where $p$ and $q$ are integers.\end{nTh}
An easy consequence is:

\begin{nTh} Any metric of the form 
\beq\label{metgen} 
g=A^2(\tht)\,d\tht^2+\sin^2\tht\,d\phi^2
\eeq
for which $A$ is given by
\[A=\frac pq+a(x)\qq\qq x=\cos\tht\]
where $p$ and $q$ are integers and $a(x)$ is an odd function is a $(p,\,q)$ Tannery metric.\end{nTh}

\nin{\bf Proof:} The rotation function, after the change of variable $\,\cos\tht=(\cos i)\cdot u$, becomes
\beq
R(g,i)=\int_{-1}^1\frac{\sin i\,(p/q+a(u\,\cos i))du}{\sqrt{1-u^2}(1-(\cos i)^2\,u^2)}\eeq
and since $a$ is odd its contribution to the integral vanishes. The remaining integral is computed from
the change of variable $u=\sin x$ by elementary methods and we get
\beq
R(g,i)=\frac pq\,\int_{-1}^1\,\frac{\sin i\,dx}{(1-\cos^2 i\,\cos^2 x)}=\frac pq\,\pi\eeq
and we conclude using Theorem 1. $\quad\Box$

Let us add the following observations:
\brm
\item For generic values of $\,(p,q)\,$ we have Tannery surfaces 
while for $p=q=1$ we have Zoll or $C_{2\pi}$-surfaces.
\item For the case of $\,g_0=g({\mb S}^2,{\rm can})$ we have $R(g_0,i)=\pi$ so that $\,p=q=1\,$, 
the simplest example of a Zoll metric for which all of the geodesics have for length $2\pi$.
\item The measure of the surface with metric (\ref{metgen}) is 
\beq\label{measure}
\mu(M,g)=2\pi\int_{-1}^{+1}\left(\frac pq+a(x)\right)dx=\frac pq\,\mu({\mb S}^2,g_0).
\eeq
\erm

Let us prove now that the two dynamical systems with a superintegrable geodesic flow, globally defined on the manifold $\,M={\mb S}^2$, which were derived in \cite{vds}, give rise to Zoll metrics.

\section{A Zoll metric on ${\mb S}^2$}
In \cite{vds} two metrics, globally defined on ${\mb S}^2$, were given respectively in Theorems 1 
and 2 of this reference. For the reader's convenience let us recall these metrics. The first one is
\beq\label{oldmet1}
g_0=\rho_0^2\,\frac{dv^2}{D_0}+\frac{4D_0}{P_0}\,d\phi^2\qq\qq v\in\,(a,1)\qq\phi\in{\mb S}^1
\eeq
where
\beq
D_0=(v-a)(1-v^2)\quad P_0=(v^2-2av+1)^2\quad Q_0=-P_0+4(a-v)D_0\quad \rho_0=\frac{Q_0}{P_0}
\eeq
and the parameter $a\in\,(-1,1)$. 

The second one is
\beq\label{oldmet2}
g=\rho^2\,\frac{dv^2}{D}+\frac{4D}{P}\,d\phi^2\qq\qq x\in\,(-1,1)\qq \phi\in{\mb S}^1
\eeq
with
\beq\left\{\barr{l}
D=(m+x)(1-x^2)\\[4mm] P=\Big(L_+(1-x^2)+2(m+x)\Big)\Big(L_-(1-x^2)+2(m+x)\Big)\\[4mm] 
Q=3x^4+4mx^3-6x^2-12mx-4m^2-1\earr\right.\qq\rho=\frac QP
\eeq
with the following restrictions on the parameters:
\[ l>-1\qq\qq m>1.\]
We will first prove:

\begin{nth} 
The metric (\ref{oldmet1}) is nothing but the limit for $l=-1$ of the metric 
(\ref{oldmet2}).
\end{nth}

\nin{\bf Proof:} Let us define the change of coordinate 
\beq
2v=(1-a)\,x+(1+a)\qq\qq v\in\,(a,1)\quad\to\quad x\in\,(-1,+1)
\eeq
then, as a consequence of the relations
\beq
D_0=\frac{(1-a)^3}{8}\,D\qq P_0=\frac{(1-a)^4}{16}\,P\qq Q_0=\frac{(1-a)^4}{16}\,Q,
\eeq
we get
\beq
g_0=\frac 2{1-a}\,g(l=-1),\eeq
concluding the proof. $\quad\Box$

So, from now on we will consider only the metric (\ref{oldmet2}) 
with the restrictions
\beq\label{restric1}
l\in\,[-1,+\nf)\qq\qq m\in\,(1,+\nf)\eeq
and we will prove:

\begin{nth} The metric (\ref{oldmet2}) subjected to the constraints (\ref{restric1}) is a 
Zoll metric on the manifold $M={\mb S}^2$.\end{nth}  

\nin{\bf Proof:} The change of coordinate \footnote{From now on we will use the notations $s=\sin\tht$ 
and $c=\cos\tht$ to shorten the formulas.}
\beq
x=\frac{H_+-H_-+2c}{H_++H_-} \qq\qq H_{\pm}(s)=\sqrt{\dst 1-\frac{(l\pm 1)}{(l+m)}\,s^2}
\eeq
implies the following relations
\beq
\frac{4D}{P}=\frac{s^2}{l+m}\qq \frac{dx}{\sqrt{D}}=-\frac 1{\sqrt{l+m}}\,\frac{d\tht}{H_+\,H_-}\qq 
-\rho=H_+\,H_--c(H_+-H_-)
\eeq
giving for the metric the final form
\beq\label{S2met1}
G=(l+m)\,g=A^2\,d\tht^2+\sin^2\tht\,d\phi^2, \qq\qq  \tht\in(0,\pi)\qq \phi\in{\mb S}^1\eeq
where 
\beq
A=1+c\left(\frac 1{H_+}-\frac 1{H_-}\right).\eeq
The function $\,A$ has the structure required by Theorem 2, with $p=q=1$, showing that $G$ is indeed 
a Zoll metric, globally defined on ${\mb S}^2$.$\quad\Box$

\nin{\bf Remarks:}
\brm
\item From relation (\ref{measure}) the measure of this surface is $\,\mu({\mb S}^2,G)=4\pi.$
\item In the special case $l=1$ the metric simplifies to
\beq
A=1+c\left(\frac 1H-1\right)\qq H=\sqrt{1-\varrho\,s^2}\qq \varrho\in\,(0,1) \eeq
while for $l=-1$ one has 
\beq
A=1+c\left(1-\frac 1H\right)\qq H=\sqrt{1+\varrho\,s^2}\qq \varrho\in\,(0,1).
\eeq
\erm
Let us examine the cubic integrals:

\begin{nth}The cubic integrals are given by 
\beq\label{0S1S21}\barr{l}
S_1=+\cos\phi\,\Pi\,(2H+\alf\,\Pf^2)+\sin\phi\,\Pf\,(2\be\,H+\ga\,\Pf^2)\\[5mm]  
S_2=-\sin\phi\,\Pi\,(2H+\alf\,\Pf^2)+\cos\phi\,\Pf\,(2\be\,H+\ga\,\Pf^2)\earr
\eeq 
where 
\beq\label{abc1}\barr{c}\dst 
\alf=\frac 1{l+m}\left(\frac{H_+-H_-+2c}{H_++H_-}-l\right)\qq \be=-\frac 1s\,(H_+-H_-+c)\\[4mm]\dst 
\ga=\frac 1{s^3}\,\Big(H_+-H_-+c(1-H_+\,H_-)\Big) \earr
\eeq
and they are constrained by
\beq\label{SSmet1}
S_1^2+S_2^2=(2H)^3+\si_1\,(2H)^2\,\Pf^2+\si_2\,(2H)\,\Pf^4+\si_3\,\Pf^6\eeq
with
\beq\label{si1}
\si_1=-\frac{3l+m}{l+m}\qq\si_2=\frac{3l^2+2lm-1}{(l+m)^2}\qq\si_3=-\frac{l^2-1}{(l+m)^2}.\eeq
\end{nth}

\nin{\bf Proof:} We have first to find the cubic integrals, taking for hamiltonian 
\beq
H=\frac 12\left(\Pi^2+\frac{\Pf^2}{s^2}\right)\qq\qq \Pi=\frac{\Pt}{A}.
\eeq
We could use the formulas given in \cite{vds} and transform them in the new $\,(\tht,\,\phi)$ 
coordinates but the computations needed are quite hairy, so we will derive them anew, 
writing them as in (\ref{0S1S21}), where the unknown functions $(\alf,\,\be,\,\ga)$ depend 
solely on $\tht$. Imposing $\,\{H,S_1\}=0$ and $\,\{H,S_2\}=0$ gives one and the same differential system
\beq\label{diffsys}
\barr{ll}(a) &\qq  s^2\,\be'=A\\[4mm] (b) &\qq s^3\,\alf'=-s\,\be\,A-c\\[4mm] (c) &\qq 
s^2\,\ga'=\alf\,A\\[4mm] (d)\dst &\qq\dst A\left(s\ga+\frac{\be}{s}\right)=-c\left(\alf+\frac 1{s^2}\right)\earr \qq\quad A=1+c\left(\frac 1{H_+}-\frac 1{H_-}\right).\eeq
Integrating (a) gives $\be$ which allows to integrate (b) giving $\alf$ which allows in turn to 
integrate (c) for $\ga$. These quadratures generate 3 unknown constants which are fixed up using (d). 
The results of these elementary computations are given in (\ref{abc1}).

It follows that
\beq\label{Scarre1}
S_1^2+S_2^2=\Pi^2\,(2H+\alf\,\Pf^2)^2+\Pf^2\,(2\,\be\,H+\ga\,\Pf^2)^2.
\eeq
Expanding, with the help of the relation $\,\dst\Pi^2=2H-\frac{\Pf^2}{s^2}$, we find 
an homogeneous polynomial of degree 3 in $2H$ and $\Pf^2$. The computation of the various 
coefficients shows that they reduce to the constants given in (\ref{si1}). $\quad\Box$

Using the initial conditions (\ref{init}) the relation (\ref{Scarre1}) becomes \footnote{From 
now on we will use the shorthand notations $\,s_0=\sin i$ and  $\,c_0=\cos i$.}
\beq\label{SScarre1}
S_1^2+S_2^2=c_0^2\,H_+^2(s_0)\,H_-^2(s_0).\eeq 
The interpretation of these cubic integrals will now be given:

\begin{nth} The conserved quantities $\,(S_1,\,S_2)$ give the parametric representation of the geodesics:
\beq\label{geod1} \left\{\barr{l}
c_0\,R(s_0)\,\sin\phi=\Pi\,(1+s_0^2\,\alf(s))\\[4mm]   c_0\,R(s_0)\,\cos\phi=-s_0\,(\be(s)+s_0^2\,\ga(s))\earr\right. 
\qq\qq \Pi=\eps\,\sqrt{1-\frac{s_0^2}{s^2}}\eeq
where the functions $\,(\alf,\,\be,\,\ga)$ are given in (\ref{abc1}) and $\eps=+1$ (resp. $\eps=-1$) 
if $\tht$ is increasing (resp. decreasing).
\end{nth}

\nin{\bf Proof:} Since $S_1$ and $S_2$ are preserved under the geodesic flow, we can compute their 
value at the starting point of the geodesic. With our choice (\ref{init}) of initial conditions we get
\beq\barr{l} 
S_1=\cos\phi\,\Pi\,(1+s_0^2\,\alf)+\sin\phi\,s_0\,(\be+s_0^2\,\ga)=0\\[4mm]
S_2=-\sin\phi\,\Pi\,(1+s_0^2\,\alf)+\cos\phi\,s_0\,(\be+s_0^2\,\ga)=-c_0\,H_+(s_0)\,H_-(s_0)\earr
\eeq
which are in agreement with (\ref{SScarre1}). These relations are easily inverted and give (\ref{geod1}), 
expressing the azimuthal angle $\phi$ parametrically in terms of the angle $\tht$. Since $\phi$ is an increasing function, using these equations we can check that for $\tht=0$ we have $\phi=0$ and for  $\tht=\pi-i$ we get $\phi=\pi$, hence $p=q=1$ in agreement with the rotation function.  $\quad\Box$

As to the embedding in ${\mb R}^3$ there is little hope to get it for generic values of the parameters 
$\,(l,\,m)$. For $l=-1$, which should be a simpler case, we may come back to the form (\ref{oldmet1}). 
Defining the cartesian coordinates as
\beq
X=A(v)\,\sin\phi\qq\qq Y=A(v)\,\cos\phi \qq\qq Z=B(v)\eeq
we get for the induced metric
\beq
g=\Big((A')^2+(B')^2\Big)\,dv^2+A^2\,d\phi^2\qq\qq v\in\,(a,1)\qq \phi\in\,{\mb S}^1.\eeq
Identifying this metric with (\ref{oldmet1}) gives for relations
\beq
A=2\frac{\sqrt{(v-a)(1-v^2)}}{v^2-2av+1}\qq (B')^2=-8\frac{(v^3-3v+2a)}{(v^2-2av+1)^3}.
\eeq
The positivity of $(B')^2$ is ensured for $\,a\in\,[0,1)$, however, since the underlying algebraic 
curve is hyperelliptic of genus $2$, its integration will be quite technical.

Let us consider now other metrics, no longer defined on a closed manifold, but rather on an orbifold.  
This weakening of the concept of manifold, introduced by Thurston, allows for a finite number of conical singularities. We will begin with the simplest example.

\section{Tannery's pear}
In \cite{vds}, Proposition 7, it is proved that the metric
\beq\label{mT}
g=\rho^2\,\frac{dv^2}{D}+\frac{4D}{P}\,d\phi^2\qq\qq v\in(-\nf,a)\qq \phi\in{\mb S}^1
\eeq
where
\beq
D=v^2(a-v)\qq P=v^2(v-2a)^2 \qq \rho=\frac{v(3v-4a)}{(v-2a)^2}\qq a>0\eeq
has a regular point for $v\to a-$ and a conical singularity for $v\to -\nf$. So it cannot be defined 
on a manifold but we will prove:

\begin{nth} The metric (\ref{mT}) is the metric of Tannery's pear defined on ${\cal T}^2$.\end{nth}

\nin{\bf Proof:} The first change of variable $\,w=1-v/a$ gives
\beq\label{Tan0}
a\,g=\frac{(1+3w)^2}{(1+w)^4}\frac{dw^2}{w}+\frac{4w}{(1+w)^2}\,d\phi^2\qq w\in (0,+\nf)\qq 
\phi\in{\mb S}^1\eeq
and it shows that we can take $a=1$. The second change of variable 
\beq
w=\frac{1+\cos\tht}{1-\cos\tht}\qq\qq w \in\,(0,+\nf)\ \to\ \tht\in\,(0,\pi)\eeq
transforms the metric into
\beq\label{Tanmet}
g=(2+\cos\tht)^2\,d\tht^2+\sin^2\tht\,d\phi^2\qq\qq \tht\in(0,\pi)\qq  \phi\in{\mb S}^1
\eeq
on which we recognize the metric on Tannery's pear \cite{Ta}. This was the first example (in 1892!) of a metric, with non-constant sectional curvature, for which the geodesics close after two turns.

We will define Tannery's orbifold ${\cal T}^2$ by the following singularity structure of the metric:
\brm
\item At the south pole $\tht\to\pi-$ we have
\beq 
g\sim d\tht^2+\sin^2\tht\,d\phi^2\qq\qq \phi\in{\mb S}^1
\eeq
showing that this point is in fact a regular point since the apparent singularity would disappear using local cartesian coordinates. 
\item However for the north pole $\tht\to 0+$ we have the conical singularity
\beq
\frac g9\sim d\tht^2+\sin^2\tht\,\left(\frac{d\phi}{3}\right)^2\eeq
precluding a manifold but allowed for an orbifold. $\quad\Box$ \erm

Let us recall some known facts about this orbifold:
\brm
\item Using (\ref{measure}) the measure of Tannery's pear is $\ \mu({\cal T}^2,g)=8\pi$.
\item Its sectional curvature 
\beq
\si({\cal T}^2,g)=\frac 2{(2+\cos\tht)^3}\eeq
is $C^{\nf}$ and positive. This implies that $G$ can be isometrically embedded in ${\mb R}^3$. If we take for 
explicit (global) embedding 
\beq
X=\sin\tht\,\cos\phi\qq Y=\sin\tht\,\sin\phi\qq Z=4\sqrt{2}\,\sin\frac{\tht}{2}
\qq \tht\in\,[0,\pi] \qq \phi\in\,{\mb S}^1\eeq
its cartesian equation is
\beq
X^2+Y^2=\frac{Z^2}{8}\left(1-\frac{Z^2}{32}\right)\qq\qq Z\in\,[0,4\sqrt{2}].\eeq
\erm
In this way the point $(X,\,Y,\,Z)=(0,0,4\sqrt{2})$ is regular while the point $(0,0,0)$ is the vertex of a cone 
with an aperture of $\,2\arctan\left(\frac 1{2\sqrt{2}}\right)$ close to $39^{\circ}$.

\begin{nth}The cubic integrals are given by 
\beq\label{0S1S22}\barr{l}
S_1=+\cos\phi\,\Pi\,(2H+\alf\,\Pf^2)+\sin\phi\,\Pf\,(2\be\,H+\ga\,\Pf^2)\\[5mm]  
S_2=-\sin\phi\,\Pi\,(2H+\alf\,\Pf^2)+\cos\phi\,\Pf\,(2\be\,H+\ga\,\Pf^2)\earr
\eeq 
where 
\beq\label{abc2} 
\alf=-\frac{(1+c)^2}{s^2}\qq\qq  \be=-\frac{(1+2c)}{s}\qq\qq \ga=\frac{(1+c)^2}{s^3}
\eeq
and they are constrained by
\beq\label{SSmet2}
S_1^2+S_2^2=(2H)^3-2\,(2H)^2\,\Pf^2+(2H)\,\Pf^4=c_0^4.\eeq
The geodesic equations are 
\beq\label{geod2}\left\{\barr{l} 
c_0^2\,\sin\phi=-\Pi\,(1+s_0^2\,\alf(s))\\[4mm]  c_0^2\,\cos\phi=s_0\,(\be(s)+s_0^2\,\ga(s)) 
\earr\right. \qq \Pi=\eps\,\sqrt{1-\frac{s_0^2}{s^2}}.\eeq
\end{nth}

\nin{\bf Proof:} The function $A=2+c$ being given, one has to integrate the differential system 
(\ref{diffsys}) as already explained in the proof of Proposition 3. Having fixed up $S_1$ and $S_2$ one 
deduces that
\beq
S_1^2+S_2^2=(2H)^3-2\,(2H)^2\,\Pf^2+(2H))\,\Pf^4=c_0^4\eeq
and upon inversion of the relations $S_1=0$ and $S_2=c_0^2$ one gets (\ref{geod2}). 
Here too we can check that for $\tht=0$ we have $\phi=0$, for $\dst\tht=\frac{\pi}{2}$ we have 
$\phi=\pi+i$ and for $\tht=\pi-i$ we get $\phi=2\pi$, hence $p=2$ and $q=1$ in agreement with the 
rotation function.  $\quad\Box$ 

\nin The equations (\ref{geod2}) for the geodesics were first given by Tannery in \cite{Ta}. 

Let us proceed with a one parametric extension of Tannery's metric.

\section{A generalization of Tannery's pear}
It was proved in Proposition 12 of \cite{vds} that the metric 
\beq\label{met2}
g=\rho^2\frac{dv^2}{D}+\frac{4D}{P}\,d\phi^2\qq\qq  v\in\,(-\nf,v_0)\qq \phi\in\,{\mb S}^1
\eeq
with
\beq
v_0>v_1\qq D=(v_0-v)(v-v_1)^2\qq P=(v-v_1)^2\,p \eeq
and
\beq
p=v^2-2(2v_0+3v_1)v+(2v_0+v_1)^2\quad \De(p)<0 \qq \rho=\frac{(v-v_1)(3v-4v_0+v_1)}{p}\eeq
Has a regular end-point for $v\to v_0-$ but  a conical singularity for $v \to -\nf$.

Let us first clean up these formulas using the change of variable
\[w=\frac{v_0-v}{v_0-v_1}\qq\qq w\in\,(0,+\nf)\]
which gives
\beq\label{metw}
(v_0-v_1)\,g=\frac{(1+3w)^2}{(w^2+2aw+1)^2}\,\frac{dw^2}{w}+\frac{4w}{w^2+2aw+1}\,d\phi^2\qq\qq a\in(-1,+1)
\eeq
so we can  set $v_0-v_1=1$ and we are left with a single parameter, namely $a$. In the 
limit $a\to 1$ we recover Tannery's metric (\ref{Tan0}).

Let us prove:

\begin{nth} The metric given by (\ref{met2}) is a $(2,1)$ Tannery metric defined on ${\cal T}^2$.\end{nth}

\nin{\bf Proof:} The change of coordinate
\beq
w=-a+\frac{(1+a)}{s^2}(1+c\,R)\qq R(s)=\sqrt{1+\vrho\,s^2}\qq \vrho=\frac{1-a}{1+a}\,\in\,(0,+\nf)
\eeq
maps  $\ w\in\,(0,+\nf)\ \to\ \tht\in\,(0,\pi)$. Using the relations
\beq\barr{c}\dst 
\frac{dw}{\sqrt{w}\,(w^2+2aw+1)}=-\sqrt{\frac 2{1+a}}\,\frac{d\tht}{1+w}\\[5mm]\dst 
\frac{2w}{w^2+2aw+1}=\frac{s^2}{1+a}\qq\qq  \frac{1+3w}{1+w}=2+\frac cR \earr\eeq 
one gets for the transformed metric
\beq\label{Tmet2}
G\equiv \frac{(1+a)}{2}\,g=A^2\,d\tht^2+s^2\,d\phi^2 \qq\qq A=2+\frac{c}{R}.
\eeq
Since $A$ has the structure required by Theorem 2, with $p=2$ and $q=1$, we conclude that it is a 
Tannery metric. The structure of the singularities for $\tht=0$ and $\tht=\pi$ agree with 
Tannery's orbifold. $\quad\Box$

Let us proceed to:

\begin{nth}The cubic integrals are given by 
\beq\barr{l}
S_1=+\cos\phi\,\Pi\,(2H+\alf\,\Pf^2)+\sin\phi\,\Pf\,(2\be\,H+\ga\,\Pf^2)\\[5mm]  
S_2=-\sin\phi\,\Pi\,(2H+\alf\,\Pf^2)+\cos\phi\,\Pf\,(2\be\,H+\ga\,\Pf^2)\earr
\eeq 
where 
\beq\label{abc3}
\alf=-\frac{1+c^2+2c\,R}{s^2}\qq\quad \be=-\frac{2c+R}{s}\qq\quad \ga=\frac{2c+(1+c^2)R}{s^3},
\eeq
and they are constrained by
\beq\label{Scarre3}
S_1^2+S_2^2=(2H)^3-(2-\vrho)(2H)^2\,\Pf^2+(1-2\vrho)(2H)\,\Pf^4+\vrho\Pf^6=c_0^4\,R^2(s_0).\eeq
The geodesics equations are 
\beq\label{geod3}\left\{\barr{l} 
c_0^2\,R(s_0)\,\sin\phi=-\Pi\,(1+s_0^2\,\alf(s))\\[4mm]   
c_0^2\,R(s_0)\,\cos\phi=s_0\,(\be(s)+s_0^2\,\ga(s))\earr\right. \qq \Pi=\eps\,\sqrt{1-\frac{s_0^2}{s^2}}
\eeq
\end{nth}

\nin{\bf Proof:} We need to integrate the differential system (\ref{diffsys}) with $\,\dst A=2+\frac cR$ 
following the same pattern explained in the proof of Proposition 3. The results are given in (\ref{abc3}). 

The proof of (\ref{Scarre3}) is similar to the one given for (\ref{SSmet1}) in Proposition 3. Using the initial conditions (\ref{init})  we have $\,S_1=0$ and $S_2=c_0^2\,R(s_0)$. Inverting 
these relations for $\sin\phi$ and $\cos\phi$ leads to the geodesics equations (\ref{geod3}). $\quad\Box$
 
Let us add the following remarks:
\brm
\item We have analyzed previously Tannery's pear for its own historical interest but in fact it appears 
as the special case $\vrho \to 0$ of the present metric (\ref{Tmet2}). 
\item The measure of this surface is still $\,\mu({\cal T}^2,G)=8\pi$.
\item Let the embedding in $\,{\mb R}^3$ be given by
\beq
x=A(w)\,\cos\phi\qq\qq y=A(w)\,\sin\phi\qq\qq z=B(w)\eeq
where we come back to the initial metric $g$ given by (\ref{metw}). It follows that we have
\[A=2\sqrt{\frac wp}\qq\qq p=w^2+2aw+1\qq a\in\,(-1,1)\]
and the function $\,B$ is given by
\beq\label{Bprime}
\left(B'\right)^2=\frac{(3p+2w+2a)(3wp+2aw+2)}{p^3}
\eeq
In the range allowed for $a$ we have $p>0$ and the resultant shows that the two polynomials in the numerator have no common zero. Hence, if $3p+2w+2a$ has real zeroes the sign of the right hand side 
will change and this happens for $\dst a\in\,(-1,-\frac{2\sqrt{2}}{3}]$, precluding any emmbedding. 
This corresponds to values of $\vrho$ larger than $33.97\ldots$. For 
$\dst a\in(-\frac{2\sqrt{2}}{3},1)$ then $3p+2w+2a>0$ and the discriminant of $3wp+2aw+2$ is strictly negative ensuring a single real root which can be checked, from Cardano formula, to be strictly 
negative and the embedding is possible. The integration of (\ref{Bprime}) will be difficult since 
the underlying algebraic curve is hyperelliptic of genus $3$.

As a side remark let us observe that the sectional curvature is strictly positive for $a \in\,(-1/3,1)$, but we have found a larger domain for $a$ in which the embedding is still possible. 
\erm

\section{A Zoll metric on Tannery's orbifold}
It was proved in Proposition 13 of \cite{vds} that the metric
\beq
g=\rho_0^2\,\frac{dv^2}{D_0}+\frac{4D_0}{P_0}\,d\phi^2\qq v\in\,(-\nf,v_0)\qq\phi\in{\mb S}^1\eeq
with
\beq
D_0=(v_0-v)[(v-v_1)^2+v_2^2]\qq v_1>0 \qq v_2\in{\mb R}\bs\{0\}\eeq
and
\beq
\rho_0=-1+\frac{4(v_0+2v_1-v)D_0}{P_0} \qq\qq  P_0=8v\,D_0+(D'_0)^2\eeq
has a regular south pole for $v\to v_0$ and a conical singularity for $v\to -\nf$.

To clean up this metric let us make the change of variable
\[v=v_0-\la\,w \qq\qq \la=\sqrt{(v_0-v_1)^2+v_2^2}\]
which yields
\beq\label{metorb3}
\la\,g=\rho^2\,\frac{dw^2}{D}+\frac{4D}{P}\,d\phi^2\qq \qq w\in\,(0,+\nf)\eeq
with
\beq\left\{\barr{l}\dst 
D=w(w^2-2aw+1)\qq\qq a=\frac{v_0-v_1}{\la}\in\,(-1,1)\\[4mm]\dst 
P=\Big(w^2+2(m+r)w+1\Big)\Big(w^2+2(m-r)w+1\Big)\qq m=2\frac{v_1}{\la} \in (0,+\nf)\\[4mm]\dst 
\rho=-1+\frac{4(w+m)D}{P}\qq\qq r=\sqrt{m^2+2am+1}.\earr\right.
\eeq
So from now on we will take $\la=1$. Let us begin with:

\begin{nth} The metric (\ref{metorb3}), still defined on ${\cal T}^2$, is a Zoll metric.
\end{nth} 

\nin{\bf Proof:} The change of coordinate
\beq
w=a+\frac{B+c\,\sqrt{2m\,B}}{s^2} 
\qq w\in\,(0,+\nf)\ \to\  \tht\in(0,\pi)\eeq 
with
\[B(s)=m\,c^2-a\,s^2+{\cal B}(s)\qq\qq {\cal B}(s)=\sqrt{(m\,c^2-a\,s^2)^2+(1-a^2)s^4}\]
and the relations
\beq
\frac{4D}{P}=\frac{s^2}{m}\qq\qq \rho^2\,\frac{dw^2}{D}=\frac 1m\,(1+c\,R)^2d\tht^2\qq
R=\sqrt{m}\,\frac{(2\,B)^{3/2}}{B^2+(1-a^2)\,s^4}
\eeq
give for the metric
\beq
G\equiv m\,g=A^2\,d\tht^2+\sin^2\tht\,d\phi^2\qq\qq A=1+c\,R.
\eeq
Since $A$ has the structure required by Theorem 2 with $p=q=1$, we get a Zoll metric and 
the singularity structure leads again to Tannery's orbifold ${\cal T}^2$. $\quad\Box$

\begin{nth}The cubic integrals are given by 
\beq\barr{l}
S_1=+\cos\phi\,\Pi\,(2H+\alf\,\Pf^2)+\sin\phi\,\Pf\,(2\be\,H+\ga\,\Pf^2)\\[5mm]  
S_2=-\sin\phi\,\Pi\,(2H+\alf\,\Pf^2)+\cos\phi\,\Pf\,(2\be\,H+\ga\,\Pf^2)\earr
\eeq 
where 
\beq\label{abc4}\barr{c}\dst 
\alf=-\frac 1{s^2}\left(\frac{{\cal B}}{m}+c\,\sqrt{\frac{2B}{m}}\right)\qq  
\be=-\frac 1s\,\left(\sqrt{\frac{2B}{m}}+c\right)\\[5mm]\dst  
\ga=\frac 1{s^3}\,\left(\sqrt{\frac{2B}{m}}+c\, \frac{{\cal B}}{m}\right) \earr
\eeq
and they are constrained by
\beq
S_1^2+S_2^2=(2H)^3+\si_1\,(2H)^2\,\Pf^2+\si_2\,(2H)\,\Pf^4+\si_3\,\Pf^6=c_0^2\,{\cal B}^2\eeq
with
\beq\label{si4}
\si_1=-\frac{3m+2a}{m}\qq\si_2=\frac{3m^2+4am+1}{m^2}\qq\si_3=-\frac{m^2+2am+1}{m^2}.\eeq
The geodesics equations are 
\beq\label{geod4}\left\{\barr{l}\dst  
c_0^2\,\frac{{\cal B}(s_0)}{m}\,\sin\phi=-\Pi\,(1+s_0^2\,\alf(s))\\[4mm]\dst    
c_0^2\,\frac{{\cal B}(s_0)}{m}\,\cos\phi=s_0\,(\be(s)+s_0^2\,\ga(s))\earr\right.\qq  \Pi=\eps\,\sqrt{1-\frac{s_0^2}{s^2}}\eeq
\end{nth}

One may wonder whether one could generalize to include in the hamiltonian some potential term 
invariant under the Killing vector $\pt_{\phi}$. This is not possible in contrast with the Koenigs  superintegrable models \cite{Ko} with quadratic integrals (special cases of Matveev and Shevchishin models) which were shown in \cite{kk} to allow for potentials.

\section{Conclusion}
As we have seen the conjecture of Matveev and Shevchishin is valid for their superintegrable model 
with cubic integrals in the momenta, be the metrics defined on manifolds or on orbifolds. It opens 
a very large field of research and of construction 
of superintegrable models with higher degrees integrals, the final target being a proof that their 
conjecture remains valid for any  such degree higher than 3. Let us observe that Kiyohara \cite{Ki} has constructed a set of {\em integrable} models with integrals of any degree greater than 3 for which all 
of their geodesics are closed. This seems to indicate that Tannery and Zoll metrics are tightly related to  
integrable and superintegrable models. 

\vspace{5mm}
\nin{\bf Acknowledgements:} We are greatly indebted to Professor David Kohel, from IML at Luminy, for 
the determination of the genera of the algebraic curves in Section 3 and in Section 5.

\end{document}